\documentclass[preprint, amssymb, amsmath, aip, pop]{revtex4-1}
\usepackage{docs}
\usepackage{bm}
\usepackage[colorlinks = true, urlcolor = blue, linkcolor = blue, citecolor = blue]{hyperref}

\usepackage{graphicx}
\usepackage{dcolumn}
\usepackage{amsfonts}
\usepackage[utf8]{inputenc}
\usepackage[usenames]{xcolor}

\usepackage{definitions}
\begin{document}
\title{Drift effects on electromagnetic geodesic acoustic modes}
\author{R. J. F. Sgalla}
\email{reneesgalla@gmail.com}
\noaffiliation
\affiliation{\mbox{\hspace{-0.177cm}\textcolor{white}{\textbf{I}}\emph{\hspace{-0.2cm}Institute of Physics, University of São Paulo, São Paulo 05508-900, Brazil}}}
\date{24 October 2014}
\begin{abstract}
A two fluid model with parallel viscosity is employed to derive the dispersion relation for electromagnetic geodesic acoustic modes (GAMs) in the presence of drift (diamagnetic) effects. Concerning the influence of the electron dynamics on the high frequency GAM, it is shown that the frequency of the electromagnetic GAM is independent of the equilibrium parallel current but, in contrast with purely electrostatic GAMs, significantly depends on the electron temperature gradient. The electromagnetic GAM may explain the discrepancy between the $f\sim 40$ kHz oscillation observed in TCABR [Yu. K. Kuznetsov \textit{et al.}, Nucl. Fusion \bf{52}, 063044 (2012)] and the former prediction for the electrostatic GAM frequency. The radial wave length associated with this oscillation, estimated presently from this analytical model, is $\lambda_r\sim 25$ cm, i. e., an order of magnitude higher than the usual value for zonal flows (ZFs).
\end{abstract}
\maketitle
\section{Introduction}
\l{I}

Zonal flows (ZFs)\cit{DiamondPPCF2005} and associated geodesic acoustic modes (GAMs)\cit{WinsorPF1968} play a crucial role in the drif wave (DW) turbulence suppression mechanism by shear flows\cit{BiglariPFB1990, HallatschekPRL2001, ItohPPCF2005}. Understanding deeply this mechanism have been challenging but may be essential to improve plasma confinement in tokamaks\cit{BiglariPFB1990, TerryRMP2000}. In addition to its relevance in the  important problem of anomalous transport, the investigation of GAMs, just like of BAEs (Beta-induced Alfvén Eigenmodes)\cit{HeidbrinkPRL1993, HeidbrinkPoP1999, ZoncaPPCF1996}, have been employed to diagnose the safety factor ($q$) and the ion temperature $(\Ti$) radial profiles. For this purpose, MHD and GAM spectroscopy techniques have been developed\cit{SharapovPLA2001, GorelenkovPoP2009, ItohPPCF2007}. 

Low frequency oscillations (4 - 40 kHz in TCABR\cit{KuznetsovNF2012}) observed mostly during the L-H transition\cit{WagnerPRL1982} have been reported in a variety of different size tokamaks\cit{KramerFleckenPRL2006, ConwayPPCF2005, McKeePPCF2006, BerkNF2006, MelnikovPPCF2006, KuznetsovNF2012}. Many of these oscillations have been recognized as ZFs and GAMs due to certain features regarding density perturbations and low poloidal modes ($m = 0, 1$)\cit{KramerFleckenPRL2006}. For more details concering experimental issues on these modes an extense review presented by Fujisawa\cit{FujisawaNF2009} can be studied.

Since the early works of Winsor\cit{WinsorPF1968} and Mikhailoviskii\cit{MikhailovskiiNF1973}, several fluid\cit{MikhailovskiiPPR1999, WatariPST2006, SmolyakovPLA2008, SgallaPLA2013} and kinetic\cit{MikhailovskiiPPR1999a, ZoncaPPCF1996, WatariPoP2005, ZoncaEL2008, ElfimovPoP2009} models have been developed to improve the theoretical prediction of the GAMs frequencies. Numerous corrections on the dynamics of these modes were included to account for realistic conditions observed in experiments and, among these corrections, those due to electromagnetic effects\cit{ZhouPoP2007, SmolyakovPPCF2008, SmolyakovNF2010, ElfimovPoP2013, LakhinPLA2014} may be essential to predict correctly not only the value of the higher frequency of GAMs, $f_{GAM}\sim \Te/\mi\rz$ (hereafter referred simply as the GAM frequency), but also the parallel current density ($\tilde{j}_\parallel$). Here $\Te$ is the electron temperature, $\mi$ is the ion mass and $\rz$ is the major radius of the tokamak. For an estimative of the radial wave length ($\lambda_r$) associated with the radial electric field, informations about electromagnetic perturbations is also necessary. The electromagnetic character of GAMs is determined by the dimensionless parameter $\K = \kr\lambda_{De}\kp c/\o$ known as the radial mode width\cit{SmolyakovNF2010}, where $\kr = 2\pi/\lambda_r$, $\kp = 1/q\rz$ and $\lambda_{De} = \sqrt{\epz\Te/ne^2}$. An alternative form to represent $\K$ is 
\bel{I0}
\K = \sqrt{\f{\te}{2}}\f{v_A}{\vthi}\f{\zc{}}{\O} = \sqrt{\f{\te}{2}}\sqrt{\f{1 + \te}{\beta}}\f{\zc{}}{\O}\co 
\ee
where $\vthi = \sqrt{2\Ti/\mi}$, $v_A = B/\sqrt{\muz n\mi}$, $\beta = 2\muz n(\Ti + \Te)/B^2$, $\O = q\rz \o/\vthi$ and $\te = \Te/\Ti$. Since electrostatic (electromagnetic) modes occurs for $\K\gg1$ ($\K\ll1$) and $\K\propto \beta^{-1/2}\lambda_r^{-1}$, electromagnetic GAMs occurs in  large beta plasmas and/or for large radial wave length perturbations. 

In high beta plasmas, such as those produced by energetic particles in NSTX\cit{GorelenkovPLA2007}, for example, electromagnetic GAMs can interact linearly with Alfvén eigenmodes (AEs), producing Beta-induced Alfvén-Acoustic Eigenmodes (BAAE)\cit{GorelenkovPLA2007, GorelenkovPoP2009}. In fact, GAMs and BAEs display some common features, like the same dispersion relation for $m = n = 0$, for instance\cit{NguyenPoP2008a, SmolyakovNF2010}. Although in low beta plasmas electrostatic GAMs are expected to be found, recently, an oscillatory mode matching GAMs' features with frequency $f\sim 40$ kHz significantly higher than that predicted theoretically for the electrostatic GAM was observed at the edge of a low beta plasma in TCABR\cit{KuznetsovNF2012}. This observation suggests that electromagnetic GAMs can occur in low beta plasmas too and, in  this case, the wave length associated with the radial perturbed electric field can be an order of magnitude higher than the usual\cit{FujisawaNF2009}. Since analytical studies on GAMs when perturbations in the magnetic field are considered in conjunction with radial gradients of density and temperature still lack in the literature, I devoted this paper to provide a study in this direction. 

In this paper, from the two fluid model an analytical expression for the GAM frequency is derived by taking into account drift effects due to density, ion and electron temperature gradients for general $\K$, thus extending the analysis for electrostatic ($\K\gg1$) and electromagnetic  ($\K\ll1$) cases. The comparison between my theoretical prediction and experimental results from TCABR\cit{KuznetsovNF2012, SeveroNF2003, NascimentoNF2005} is also provided presently. Despite the fact that the derivation of the dispersion relation for ZFs and GAMs in the leading order involves only first harmonics, second harmonics of the magnetic potential are important to determine the parallel current density\cit{ZhouPoP2007, ElfimovPoP2013} and, consequently, the value of $\K$, which allows to compute the magnetic potential, may be used to guide future experiments and/or to compare with them.

The first prediction of the GAM frequency\cit{WinsorPF1968}, $\o_{GAM}^2 = \gamma(2 + q^{-2})(\Ti + \Te)/\mi\rzt$, which came from the ideal magnetohydrodynamics (MHD) theory, was in disagreement with the kinetic result\cit{LebedevPoP1996, ZoncaPPCF1996, ElfimovPoP2010}. For such an agreement within the fluid theory, not only different adiabatic indexes for ions and electrons need to be considered ($\gamma_e\neq \gamma_i = \gamma$), but also the effect of parallel viscosity or, equivalently, pressure anisotropy ($p_\perp\neq p_\parallel$). In Refs. \cit{SmolyakovPLA2008, SgallaPLA2013}, $\gamma_i = 5/3$, $\gamma_e = 1$ and ion parallel viscosity ($\pi_{i\parallel}$) are considered and, apart from correction of $\Or{q^{-2}}$ and related Landau damping\cit{ZoncaPPCF1996, SugamaJPP2006, ZoncaEL2008, ElfimovPoP2010}, the kinetic result was recovered, viz., $\o_{GAM}^2 = 2(7\Ti/4 + \Te)/\mi\rzt$. Later, drift effects due to density and ion temperature gradients were included to the dynamics of GAMs in the electrostatic limit ($\K\rar\infty$) and high safety factor approximation ($q\rar\infty$)\cit{SgallaPLA2013} leading to the following frequencies: 
\bel{I2}
\O_{GAM\pm}^2 = \f{(\ogzt + \Ose^2)}{2}\bc{1 \pm \sqrt{1 + \f{(4\etai - 3)q^2\Ose^2}{(\ogzt + \Ose^2)^2}}}\co 
\ee
where $\O_{GAM} = q\rz\o_{GAM}/\vthi$, $\ogzt = (7/4 + \te)q^2$, $\Ose = -q\rv{i}\rz/2r\Ln$ is ratio between the drift frequency and the transit frequency ($\vthi/q\rz$), $\Ln = (\partial \ln n/\partial r)^{-1}$ is the radial density length scale, $\etai = \Lti/\Ln$ and $\Lti = (\partial \ln \Ti/\partial r)^{-1}$. In the electrostatic limit the GAMs frequencies are independent of the electron temperature gradient. The electromagnetic GAMs, on the other hand, as shown in this paper, are strongly influenced by the electron temperature gradients but not for the electron parallel velocity. Additional corrections of $\Or{q^{-2}}$, which are important mainly for the low frequency GAM ($\O_{GAM-}$), and ion Landau damping on GAMs and on the ion sound mode were investigated later (but in the electrostatic limit too) using the gyrokinetic model\cit{SgallaEPS2013}.

This paper is organized as follows. In section \r{M} we present the two fluid model considering  ions in the fluid regime ($\gamma_i = 5/3$) and electrons in the adiabatic and isotherm regime ($\gamma_e = 1$). Then, in the same section, the set of two fluid equations is solved to obtain the dispersion relation for GAMs and the parallel current ($\jpt{}$). The asymptotic analysis of the dispersion relation in the predominantly electrostatic and electromagnetic cases and the comparison between the theoretical prediction of the GAM frequency obtained in this paper and  the experimental value observed in TCABR is shown in section \r{An}. A summary and discussions follows in the last section \r{Co} and, finally, in appendix \r{A}, useful relations are provided.

\section{Model}
\l{M}

\subsection{Two fluid equations with parallel viscosity}
\l{MA}

I begin by considering the Braginskii's equations \cit{BraginskiiRPP1965} without taking into account heat flux and dissipative terms due to collisions but, instead, we consider parallel viscosity ($p_\perp\neq p_\parallel$). Ion heat flux was included in the investigation of equilibrium rotation (poloidal and toroidal) effects on GAMs and ZF and, in the leading order, it has been found that the GAM frequency remains unchanged\cit{IlgisonisPPCF2011, ElfimovPPCF2011}. In the ZF dynamics, however, heat flux must be included, at least for equilibrium with poloidal rotation and isotherm magnetic surfaces\cit{ElfimovPPCF2011}. 

The conservation of mass, momentum and energy laws are then expressed through the following  equations:
\bel{f1}
\dts{\nh{\alpha}} + \nh{\alpha}\di{\vbh{\alpha}} = 0\co
\ee
\bel{f2}
\ms\nth{\alpha}\dts{\vbth{\alpha}} + \nb\ph{\alpha} + \di{\pih{\alpha}} - \es\nh{\alpha}(\eb + \vbh{\alpha}\t\bb) = 0\co
\ee
\bel{f3}
\dts{\ph{\alpha}} + \gas\ph{\alpha}\di{\vbh{\alpha}} + (\gas -1)\pih{\alpha}:\nb\vbh{\alpha} = 0
\ee
where $\alpha$ labels for ions (i) and electrons (e), $n$, $p$ and $\vb$ stands for density, pressure and velocity, $d_\alpha/dt = \partial/\partial t + \vbh{\alpha}\cd\nb$ is the convective derivative and the ion and electron adiabatic indexes are $\gamma_i = 5/3$ and $\gamma_e = 1$, repectively. The equation describing the evolution of the viscosity tensor\cit{Mikhailovskii1BPP984, SmolyakovCJP1998}, valid for a general curvilinear magnetic field, is  
\bel{f4}
\begin{array}{l}
\ds\dts{\pih{\alpha}} + \pih{\alpha}\di{\vbh{\alpha}} + \bc{\pih{\alpha}\cd\nb\vbh{\alpha} + (\pih{\alpha}\cd\nb\vbh{\alpha})^{T} - \f23\pih{\alpha}:\nb\vbh{\alpha}\it} \\
\\
\ds\ph{\alpha}\bc{\nb\vbh{\alpha} + (\nb\vbh{\alpha})^{T} - \f23\di{\vbh{\alpha}}\it} +  \ocs(\bh\t\pih{\alpha} - \pih{\alpha}\t\bh) = 0\co
\end{array}
\ee
where higher momentums of the distribution function were neglected. Although the gyroviscous contribution of the viscosity tensor is essential to account for finite Larmour radius effects (FLR)\cit{PogutseJPP1998}, which must be considered in the analysis of eigenmode structure of GAMs\cit{ItohPPCF2005}, for example, we only include the parallel contribution, \mbox{$\pibc{\parallel}^{(i)} = (p_\parallel - p_\perp)(\bh\bh - \it/3)$} due to pressure anisotropy of ions\cit{SmolyakovPLA2008}, leaving the former issue for a future paper.

If one computes the cross product of Eq. \re{f2} with $\bb$, the general expression for the velocity is obtained, i. e., 
\bel{fvel}
\vbh{\alpha} = \f{\bb\cd\vbh{\alpha}}{B^2}\bb + \f{\eb\t\bb}{B^2} - \f{\nb\ph{\alpha}\t\bb}{\es\nh{\alpha}B^2} - \f{\di{\pih{\alpha}}\t\bb}{\es\nh{\alpha}B^2} + \f{\ms}{\es B^2}\bb\t\dts{\vbh{\alpha}}\co
\ee
but, as we consider the dynamics of GAMs, the $\eb\t\bb$-drift flow, as well as the parallel flow, is  dominant. The MHD order, $\vbh{\alpha} = \vph{\alpha}\bh + \vbE + \Or{\delta\vthi}$, is adopted in this paper as it is customary in most studies on the dynamics of GAMs\cit{HazeltineB1992, WinsorPF1968, SmolyakovPLA2008}. Here, $\bh = \bb/B$, $\bf{v}_E = \eb\t\bb/B^2\sim \vthi$, $\delta = \rv{i}/L_\perp\ll1$, $\rv{i} = \vthi/\oci$ is the ion Larmour radius, $\oci = eB/\mi$ is the ion cyclotron frequency and $L_\perp$ is the smallest perpendicular length scale of the macroscopic quantities of the plasma. 

Note, however, in the analysis of the current density, $\bf{j} = e(\nh{i}\vbh{i} - \nh{e}\vbh{e})$, that since the $\eb\t\bb$-drift flow is the same for ions and electrons, one must take the next order term to compute the perpendicular current, 
\bel{fjper}
\bf{j}_\perp = \f{\bh\t\nb\ph{i}}{B} + \f{\bh\t\nb\ph{e}}{B} +  \f{\bh\t\di{\pih{i}}}{B} + e\nh{i}\f{\bh}{\oci}\t\dt{\bf{v}_E}.
\ee
\subsection{Ion dynamics}
\l{MC}

Using ``$\widetilde{\hspace{0.4cm}}$'' to represent temporal perturbations, in eqs \re{f1}--\re{f4} we perform the following substitutions: $\nh{i} = \nz + \nth{i}$, $\vbh{i} = \vbE$, $\ph{i} = \nz\Ti + \ptc{}^{(i)}$ and $\pih{i} = \tilde{\boldsymbol{\pi}}^{(i)}$, where the perturbed ion parallel velocity, which is responsible for corrections of $\Or{\cs/q^2\rz}$ in the GAM frequency, has been neglect based on the fact that in most situations GAM are detected at positions in which $q$ is high. We do not consider equilibrium rotation for ions either, but, in the next subsection, the existence of an equilibrium electron parallel velocity is assumed. Simultaneous poloidal and toroidal equilibrium rotation effects on ZFs and GAMs have been investigated in the framework of the one fluid MHD theory\cit{IlgisonisPPCF2011, ElfimovPPCF2011}.

After performing the linearization procedure, eqs. \re{f1}, \re{f3} and \re{f4} becomes
\bel{MI1}
\dpt{\ntc{}^{(i)}} + \vbE\cd\nb\nz + \nz\di{\vbE}= 0\co
\ee
\bel{MI2}
\dpt{\ptc{}^{(i)}} + \vbE\cd\nb(\nz\Ti) + \gamma_i\nz\Ti\di{\vbE} = 0\co
\ee
\bel{MI3}
\dpt{\tilde{\pi}_\parallel^{(i)}} + \nz\Ti\bh\bh:\bc{\nb\vbE  + (\nb\vbE)^T - \f23(\di{\vbE})\it} = 0\co
\ee
and the parallel component of Eq. \re{f4} leads to the following equation:
\bel{MI4}
\mi\nz\dpt{\vpth{i}} + \np\pth{i} + \npt(\nz\Ti) + \bh\cd\di{\pith{i}_\parallel} - e\nz\Ept = 0\co
\ee
where $\np = \bh\cd\nb$, $\npt = (\bbt/B)\cd\nb$, $\bbt = \curl{(\Aptc{}\bh)}$ and $\Ept = -\np\Phitc{} - \dpth{\Apt}$.

The perturbed quantities, when expressed in the quasi-toroidal coordinate system characterized by the radial position (r), poloidal angle ($\theta$) and toroidal angles $(\phi)$, are considered to be of the form: $\tilde{x} = \sum_m\tilde{x}_m(r)\exp{[i(m\theta - \o t)]}$. Derivatives with respect to the radial position are computed via eikonal approximation for perturbed quantities \mbox{($\partial\tilde{x}/\partial r \rar \kr\tilde{x}$, $\kr\gg r^{-1}$)} and via density and temperature spatial length scales for equilibrium quantities, viz., $(\partial\ln\nz/\partial r)^{-1} = \Ln$ and $(\partial \ln T_{i, e}/dr)^{-1} = L_{T_{i,e}} = \Ln/\eta_{i,e}$.

By introducing the following perturbed normalized quantities, 
\begin{widetext}
\bel{MNP}
\Pbc{m} = \f{e\Phitc{m}}{\Ti}\co \Abc{m} = \f{e\vthi\Aptc{m}}{\Ti}\co 
\ncc{\alpha}{m} = \f{\ntc{m}^{(\alpha)}}{\nz}\co \pcc{\alpha}{m} = \f{\ptc{m}^{(\alpha)}}{\nz\Ti}\co \vcc{\alpha}{m} = \f{\vptc{m}^{(\alpha)}}{\vthi}\co
\ee
\end{widetext}
and the normalized frequencies, 
\bel{MNO}
\O = \f{q\rz\o}{\vthi}\co \Ose = \f{-q\rz\Te}{eBr\Ln\vthi} \co
\ee
after the use of relations \re{A1}, \re{A2} and \re{A3a} in eqs. \re{MI1}--\re{MI3}, it follows that 
\bel{es1}
\ncc{i}{\pm1} = \pm \bp{\f i2\f{q\zc{}}{\O}\Pbc{0} + \inv\te\f{\Ose}{\O}\Pbc{\pm1}} \co
\ee
\bel{es2}
\pcc{i}{\pm1} = \pm \bc{\gamma_i\f i2\f{q\zc{}}{\O}\Pbc{0} + \f{(1 + \etai)}{\te}\f{\Ose}{\O}\Pbc{\pm1}}\co \picc{i}{\pm1} = \pm \f i6\f{q\zc{}}{\O}\Pbc{0} .
\ee
From the solution of \re{MI4},
\bel{es4}
\vcc{i}{\pm1} = \pm\f{\pcc{i}{\pm1} + \picc{i}{\pm1} + \Pbc{\pm1}}{2\O} \mp \bc{1 + \f{(1 + \etai)}{\te}\f{\Ose}{\O}}\Abc{\pm1}\co
\ee
we conclude that $\vcc{i}{\pm1}\sim \ncc{i}{\pm1}/\O$ and since in most situation $\O_{GAM}\propto q\gg1$,  the term \mbox{$\nz\di{(\vph{i}\bh)}$} can be neglected in a first approximation. 
\subsection{Electron dynamics}
\l{MB}

First we consider the electron dynamics in equilibrium, where the magnetic field and current density for an axisymmetric system is given by
\bel{eqB}
\bb = F\nb\phi + \nb\phi\t\nb\Psi\co 
\ee
\bel{eqJ}
\jb = \muz^{-1}\curl\bb = \muz^{-1}(R^2\Delta^*\Psi\nb\phi - \nb\phi\t\nb F).
\ee
Here, $\Psi$ is the magnetic surface, $F = F(\Psi)$ is a flux function related to the toroidal magnetic field (note that the equality is satisfied in the absence of equilibrium poloidal rotation\cit{IlgisonisPPCF2011}) and $\Delta^*\Psi = \di{(\nb\Psi/R^2)}$ is the Shafranov operator. Based on the argument that the electron inertia is much lesser than the ion's, we assume that the parallel current density is due to the electron motion exclusively, i. e., $J_\parallel = -e\nz\vze$, where $\vze$ is the electron parallel equilibrium velocity.

If we take the parallel component of \re{eqJ}, for a high aspect ratio tokamaks ($\rz\gg r$) with circular magnetic surfaces ($\Psi\approx \Psi(r)$), the following equation is obtained
\bel{eqJ1}
\inv{\rz}\bp{\dprt{\Psi} - 2\inv{\rz}\dpr{\Psi}\ct} = -\muz e\nz\vze\co
\ee
where, the radial derivatives of the magnetic surfaces, in terms of the safety factor, \mbox{$q = \bb\cd\nb\phi/\bb\cd\nb\theta$}, and the magnetic shear, \mbox{$s = (r/q)dq/dr$}, can be computed as 
\bel{eqJ2}
\dpr\Psi\approx \f{rF}{q\rz}\co \dprt{\Psi} = \f{(1 - s)}{r}\dpr\Psi.
\ee
It follows that the electron equilibrium parallel flow can be represented in the leading order by
\bel{vpar}
\vze =   (s -1)(1 + \te)\f{\rv{i}}{\beta}\f{\vthi}{q\rz}.
\ee

In contrast to the ion equations, \re{MI1}--\re{MI4}, the perturbed parallel velocity for electrons must be taken into account in the development of the electron equations. In fact, for a correct description of the GAM dynamics, it is reasonable to consider $|\vpth{e}|\gg|\vbE|\gg |\vpth{i}|$. Consequently, in the leading order, the following equations must be solved:
\bel{E1}
\dpt{\ntc{}^{(e)}} + \vze\bh\cd\nb\ntc{}^{(e)} +  \nz\di{(\vptc{}^{(e)}\bh)} = 0\co
\ee
\bel{E2}
\dpt{\ptc{}^{(e)}} + \vze\bh\cd\nb\ptc{}^{(e)} + \nz\Te\di{(\vpth{e}\bh)} = 0\co
\ee
\bel{E3}
\np\pth{e} + \npt(\nz\Te) + e\nz\Ept = 0.
\ee
The correspondent solutions are:
\bel{ees1}
\vcc{e}{\pm1} = \pm \bp{\O \mp \f{\vze}{\vthi}}\ncc{e}{\pm1}\co \ncc{e}{\pm1} = \f{\pcc{e}{\pm1}}{\te}\co
\ee
\bel{ees2}
\pcc{e}{\pm1} = \Pbc{\pm1} \mp \bc{1 \mp (1 + \etae)\f{\Ose}{\O}}\O\Abc{\pm1}.
\ee

\subsection{Couple between ion and electron dynamics}
\l{ME}

As perturbations of the magnetic field is considered, the electromagnetic potential ($\Apt$) can be related to the electrostatic  potential ($\Phitc{}$). For that purpose, we consider the parallel component of the Amp\`{e}re's law,
\bel{co1}
\bh\cd(\curl\bbt) = \muz\jtc{\parallel}\co
\ee
where $\jtc{\parallel}\approx -e\nz\vthi(\vcc{e}{} + \vze\ncc{e}{}/\vthi)$ can be computed from \re{ees1}--\re{ees2} and, in terms of the magnetic potential, the LHS of \re{co1} is shown in \re{A12}. After some algebraic manipulation the relation between the electrostatic and electromagnetic potentials are provided:
\bel{co2}
\Pbc{\pm1} = \pm \bc{1 - \K^2 \mp (1 + \etae)\f{\Ose}{\O}}\O\Abc{\pm1}\co \K^2 = \f{(1 + \te)\te}{2}\f{\zc{2}}{\beta\O^2}
\ee
At this point, it should be noted that $\jtc{\parallel}$ is independent of $\vze$, and the same occurs to the potentials. 

Another constraint necessary for further analytical development is the quasi-neutrality condition,
\bel{co3}
e(\ncc{i}{\pm1} - \ncc{e}{\pm1}) = 0\co
\ee
which, when \re{es1}, \re{ees1}, \re{ees2} and \re{co2} are invoked, yields 
\bel{sA}
\Abc{\pm1} = - \f i2 \f{\te q\zc{}}{\D{\pm}}\f{\Pbc{0}}{\O^2}\co \D{\pm} = \K^2 - (1 + \etae)\f{\Ose^2}{\O^2} \pm (1 - \K^2)\f{\Ose}{\O}
\ee
\bel{spie}
\pcc{i}{\pm1} = \pm \bp{\gamma_i\f i2\f{q\zc{}}{\O}\Pbc{0} + \f{(1 + \etai)}{\te}\f{\Ose}{\O}\Pbc{\pm1}}\co  \pcc{e}{\pm1} = \mp \K^2\O\Abc{\pm1}.
\ee
Note that in the electrostatic limit ($\K\rar \infty$) eqs. \re{co2} and \re{spie} are confirmed by eqs. (13), (15) and (16) of Ref. \cit{SgallaPLA2013}.

For further analysis, it is convenient to use the trigonometric form of the perturbations, which is derived from the relations: \mbox{$\tilde{X}_{ms} = -i(\tilde{X}_{m} - \tilde{X}_{-m})$} and \mbox{$\tilde{X}_{mc} = \tilde{X}_{m} + \tilde{X}_{-m}$} for $m = 1, 2, 3, ...$. In the next step, the quantity defined as \mbox{$\P_m = \pcc{i}{m} + \pcc{e}{m} + \picc{i}{m}/4$} is properly employed to obtain the dispersion relation and to compute the second harmonic contributions from the magnetic potential. Note that 
\bel{spts}
\f{\P_s}{q\zc{}} = \bc{\f{\ogzt}{q^2} + \f{(1 + \te + \etai)}{\D{+}\D{-}} \bp{ (1 - \K^2)^2 + (1 + \etae)\K^2 - (1 + \etae)^2\f{\Ose^2}{\O^2}}\f{\Ose^2}{\O^2}}\f{\Pbc{0}}{\O}\co
\ee
\bel{sptc}
\f{\P_c}{q\zc{}} = -i(1 + \te + \etai)\f{(1 - \K^2)\K^2}{\D{+}\D{-}}\f{\Ose}{\O}\f{\Pbc{0}}{\O}
\ee

To derive the dispersion relation we use the quasi-neutrality condition again, but, this time, in the form: $\di{\jbtc{}} = 0$. After integrating this term in the plasma volume and using the Divergence's theorem, considering, for this computation, $d\bf{S} = \er r\rz(1 + \ep\ct)d\theta d\phi$ as the surface element, we obtain the following relation:
\bel{RD0}
\oint d\theta(1 + 2\ep\ct)(\jtc{I_r}^{(i)} + \jtc{p_r}^{(i)} + \jtc{\pi_r}^{(i)} + \jtc{p_r}^{(e)}) = 0\co
\ee
which, with the substitution of \re{B3}--\re{B5} in it, yields
\bel{RD1}
\oint d\theta\bc{\f{\O}{q^2}\left(q\zc{}\Pbc{0}\right) + 2\bp{\dptheta{\pcc{i}{}} + \dptheta{\pcc{e}{}}}\ct + \f32\picc{i}{}\st - \dptheta{\picc{i}{}}\ct} = 0.\q
\ee

Since $\ep\ll1$ is considered, in the leading order, the dispersion relation takes the form
\bel{RD2}
-\O\f{\zc{}}{q}\Pbc{0} + \P_s = 0\co
\ee
or, equivalently, by recalling eqs. \re{spts} and \re{RD2}, 
\bel{RD3}
\f{\O^2}{\ogzt} = 1 + \f{(1 + \te + \etai)}{(7/4 + \te)}\f{\left[(1 - \K^2)^2 + (1 + \etae)\K^2 - (1 + \etae)^2\Ose^2/\O^2\right]}{\left[\K^2 - (1 + \etae)\Ose^2/\O^2\right]^2 - (1 - \K^2)^2\Ose^2/\O^2}\f{\Ose^2}{\O^2}.
\ee
In this section we are concerned only with the analysis of the asymptotic values of the GAM frequency in the electrostatic and electromagnetic limits, viz.,  
\bel{Pslim}
\O_{GAM}^2 = \left\{\begin{array}{ll}
\ds \bp{\f74 + \te}q^2 + \f{(1 + \te + \etai)q^2}{1 - \Ose^2/\O_{GAM}^2}\f{\Ose^2}{\O_{GAM}^2}\co &  \K\rar\infty\\
\\
\ds \bp{\f34 - \etai}q^2\co & \K\rar 0
\end{array}\right.\co
\ee
leaving other cases to be studied in the next section. 

The second harmonic contribution is obtained from the equations 
\bel{}
\oint d\theta \sin(2\theta)\di{\jbtc{}} = 0\co\oint d\theta \cos(2\theta)\di{\jbtc{}} = 0\co 
\ee
which can be computed using relations \re{A7}, \re{A8} and \re{A12} and yield 
\bel{A2sc}
\Abc{2s} = - \f i4\f{\te q\zc{}}{\K^2\O^2}\tilde{\mathcal{P}}_s\co \Abc{2c} =  - \f i4\f{\te q\zc{}}{\K^2\O^2}\tilde{\mathcal{P}}_c.
\ee

Since $\Abc{2s} \propto \P_s$ and $\P_s\propto \O_{GAM}^2$, according to \re{RD2}, it can be noted that the electromagnetic contribution from the vector potential $\Apt{}$ causes a modification in the parallel current, which, in turn, exerts influence on the dynamics of the GAM's mechanism and, hence, a modification in the value of the GAM frequency is caused.

\section{Analysis of results from TCABR and derivation of the analytical expression for the GAM frequency}
\l{An}

In a recent experiment executed in TCABR by means of multi-pin Languimir probes an oscillation with frequency $f_{exp}\sim40$ kHz matching GAMs features  was observed\cit{KuznetsovNF2012}. This value, however, is about 20\% higher than the theoretical prediction for the electrostatic GAM frequency at the radial position in which $\Te\sim 30$ eV: $f_{teo}^{GAM}\sim 34$ kHz (considering ion temperature in the computation). Taking into account electromagnetic corrections, the analysis proposed in this paper may reconcile this discrepancy. According to previous experiments performed in TCABR under similar conditions\cit{SeveroNF2003, NascimentoNF2005, KuznetsovNF2012}, for the present analysis, the following density and temperatures radial profiles are considered: $\nz = N_0(1 - r^2/a^2)^{0.6}$, $\Te = T_{e0}(1 - r^2/a^2)^{1.5} + T_{ea}$ and $\Ti\approx T_{i0}(1 - r^2/a^2)+ T_{ia}$, where $T_{ea}\sim T_{ia}\sim 15$ eV is the electron/ion temperature at $r = a = 18$ cm, $T_{e0} = 400-600$ eV and $T_{i0} = 150-200$ eV are the temperatures at the center ($r = 0$) and $N_0 = 1-3\cd 10^{19}\text{m}^{-3}$. These radial profiles are not exact but can provide an estimative for the radial localization of the GAM in TCABR: $r/a\sim 0.94$. An indication that the high frequency GAM is localized close to the last magnetic surface ($\Psi_{95}$), as in T-10 tokamak\cit{MelnikovPPCF2006}, is observed in TCABR too.

To compare analytical and experimental predictions of the GAM frequency in Fig \r{f} \mbox{$\P_s/\P_{s_{\text{ref}}}$} is plotted as a function of $\K$, where \mbox{$\P_{s_{\text{ref}}}   = \lim_{\substack{\K\rar\infty \\ \Ose\rar 0}} \P_s$} is a reference term proportional to the diamagnetic contribution of the current density in the electrostatic limit when drift effects are not considered. According to \re{spts} and \re{RD2}, \mbox{$\P_s/\P_{s_{\text{ref}}}$} represents the GAM frequency normalized by its reference value, $(7/4 + \te)^{1/2}\vthi/\rz$. In Fig. \r{f}, the upper and lower dashed curves stand for the experimental and theoretical (previously obtained)  values of the  GAM frequency normalized  by the reference value mentioned before and the solid (doted) curve represents the normalized GAM frequency computed from Eq. \re{spts} for $\Ose/\O_{GAM} = 0.03$ ($\Ose/\O_{GAM} = 0.35$).  With respect to temperature gradients, in Fig. \r{f}, two scenarios are considered: $\etai = \etae = 1$ (left) and $\etai = \etae = 2$ (right). 
\begin{figure}[h!]
\centering
\includegraphics[scale = 0.43]{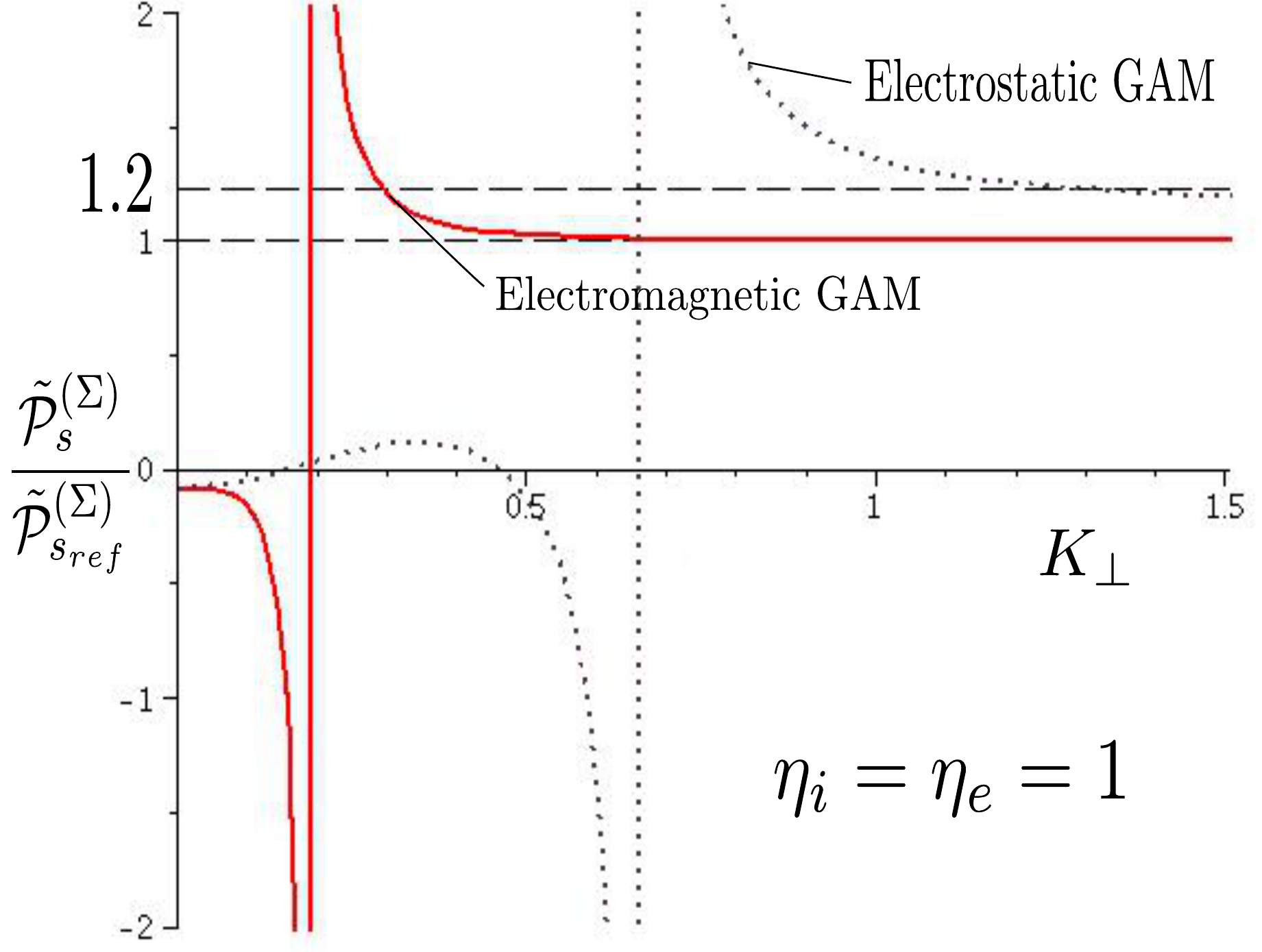}
\includegraphics[scale = 0.44]{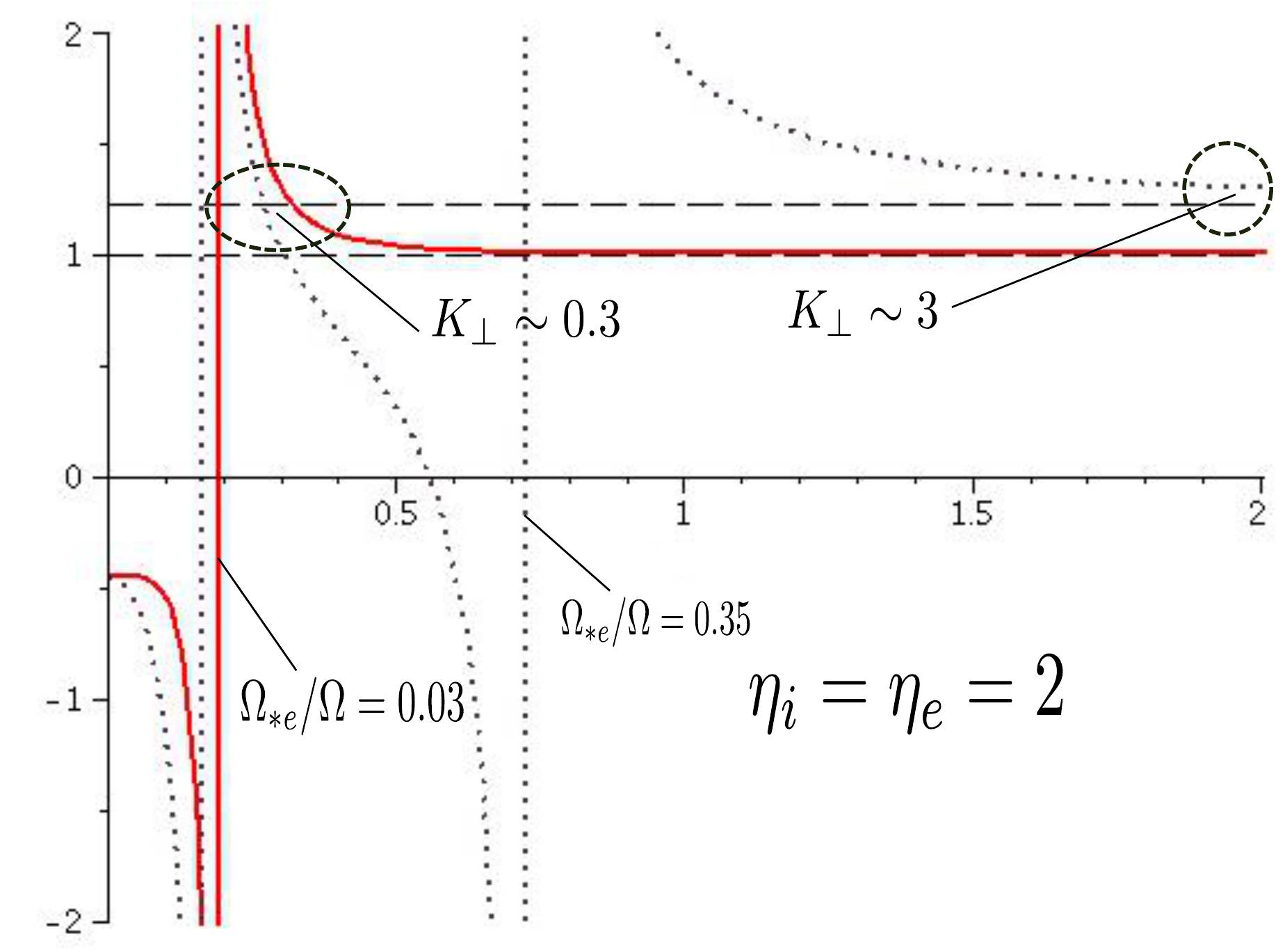}
\caption{\l{f}\mbox{$\P_s/\P_{s_{\text{ref}}}$} vs $\K$ for \mbox{$\Ose/\O = 0.03$} (solid curve) and \mbox{$\Ose/\O = 0.35$} (doted curve) and for \mbox{$\etai = \etae = 1$} (left) and $\etai = \etae = 2$ (right). The upper dashed line is the experimental value of the GAM frequency normalized by the reference value, $(7/4 + \te)q$.}
\end{figure}

From the analysis of Fig. \r{f} it is possible to note that for small gradients (probably during the L mode), $\etai = \etae = 1$ and $\Ose/\O\sim 0.03$, a crude estimative provides $\K\sim 0.3$ for the high frequency GAM generated in TCABR, indicating that it has an electromagnetic character. The correspondent radial wave length, estimated from Eq. \re{I0} under the experimental conditions described above, is $\lambda_r\sim 25$ cm, i.e., an order of magnitude higher than the usual values for ZFs and GAMs\cit{FujisawaNF2009}. For large density gradients ($\Ose/\O\sim 0.35$), possibly close to the transport barriers location, electrostatic and electromagnetic GAM can be excited, depending on the temperature gradients. 

A simple expression for the GAM frequency consistent with the experimental result from TCABR can be derived for the following specifics cases: $\K \sim 0.3$ and $\Ose/\O \sim 0.03$ (electromagnetic GAM) and $\K\sim3$ and $\Ose/\O = 0.35$ (electrostatic GAM). For this derivation, it is convenient to define the parameter \mbox{$\Delta = \Ose/\K\O\sim 0.1$}, which can be used to obtain a simpler expression for the dispersion relation in comparison with Eq. \re{RD3}, viz., 
\bel{RD4}
\f{\O^2}{\ogzt} = 1 + \f{\oett}{\ogzt}\f{\left[(1 - \K^2)^2 + (1 + \etae)\K^2 - (1 + \etae)^2\Ose^2/\O^2\right]}{\left[\K^2 - (1 + \etae)\Ose^2/\O^2\right]^2 - (1 - \K^2)^2\Ose^2/\O^2}\f{\Ose^2}{\O^2}\approx\nnb
\nnb
1 + \left\{\begin{array}{ll}
\ds \f{\Delta^2/\K^2}{1 - \Delta^2/\K^2}\bp{1 + (\etae - 1)\K^2 - (1 + \etae)^2\K^2\Delta^2}\f{\oett}{\ogzt}\co &\K\ll1\\
\\
\ds  \f{(1 + \etae)\K^2\Delta^2}{1 - (1 + \etae)\K^2\Delta^2}\f{\oett}{\ogzt}\co & \K\approx 1\\
\\
\ds \f{\K^2\Delta^2}{1 - \K^2\Delta^2}\bp{1 + \f{\etae - 1}{\K^2} - (1 + \etae)^2\f{\Delta^2}{\K^2}}\f{\oett}{\ogzt}\co &\K\gg1\\
\end{array}\right.\co
\ee
where $\oett = (1 + \te + \etai)q^2$.

The simplest case to analyse is for $\K\approx 1$ yielding two solutions, 
\bel{Sk1+}
\O_{GAM+}^2 = \ogzt + (1 + \etae)(1 + \te + \etai)q^2\f{\Ose^2}{\ogzt}\co \ogzt = \bp{\f74 + \te}q^2\co
\ee
\bel{Sk1-}
\O_{GAM-}^2 = (1 + \etae)q^2\bp{\f34 - \etai}\f{\Ose^2}{\ogzt}\co\nnb
\ee
which agree with the result obtained for electrostatic GAMs in the limit \mbox{$q\rar \infty$} and \mbox{$\etae = 0$}\cit{SgallaPLA2013, ZoncaPPCF1996, ZoncaNF2009}. Note that when  magnetic perturbations are considered in the dynamics of GAMs the effect of electron temperature gradient is enhanced. This occurs because the electron motion produces magnetic perturbations that modify the parallel current and, consequently, the diamagnetic current, which is proportional to the GAM frequency.  

In the electrostatic limit ($\K\gg1$) the solutions are
\bel{Skg1+}
\O_{GAM+}^2 = \ogzt + (1 + \te + \etai)q^2\bc{1 +  \f{2\beta(\etae - 1)\ogzt}{\te(\te + 1)\zc{2}}            }\f{\Ose^2}{\ogzt}\co
\ee
\bel{Skg1-}
\O_{GAM-}^2 = q^2\bp{\f34 - \etai}\f{\Ose^2}{\ogzt}.
\ee

The last and more important analysis is for the electromagnetic GAM ($\K\ll1$), which yields 
\bel{Skl1+}
\O_{GAM+}^2 = \ogzt + (1 + \te + \etai)q^2\bc{1 + \f{(\etae - 1)(\te + 1)}{7/2 + 2\te}\f{\te\zc{2}}{q^2\beta}}\f{\Ose^2}{\ogzt}\co 
\ee
\bel{Skl1-}
\O_{GAM-}^2\propto  \inv4\f{(\te + 1)^2}{\Ose^2}\f{\te^2\zc{4}}{\beta^2}.
\ee
Note from Eqs.\re{Sk1+}, \re{Skg1+} and \re{Skl1+} that in all cases the GAM frequency is increases with $\etae$.

\section{Summary and discussion}
\l{Co}

In this paper, asymptotic analytical expressions for the frequency of GAMs are derived from a two fluid model that considers parallel viscosity due to pressure anisotropy \mbox{($p_\perp\neq p_\parallel$)}. Consequently, ions (electrons) are assumed to be in the fluid (adiabatic and isotherm) regime(s), i.e., $\gamma_i = 5/3$ and $\gamma_e = 1$ are considered. The influence of the electron dynamics on these modes are explored in two ways. First, due to their small mass, in equilibrium, the electrons are assumed to have parallel velocity (with respect to the equilibrium magnetic field) and, thus,  an equilibrium parallel, or anti-parallel, current, depending on the value of the magnetic shear ($s = r\partial\ln q/\partial r$), according with Eq. \re{vpar}, is produced. Second, since a magnetic perturbation (mainly in the perpendicular direction) is considered, the coupling of the perturbed vector potential ($\Apt{}$) with the electron density and temperature gradients are taken into account in Eq. \re{E3}. For $q\gg1$, as considered presently, $|\vpth{e}{}|\ll |\vbE| \ll |\vpth{i}{}|$, and, although $\vpth{e}{}$ depends on the equilibrium parallel velocity (Doppler effect) as shown in Eq. \re{ees1}, the electron pressure, which is used to compute the frequency of GAMs, does not. Consequently, the GAM frequency is not modified by the electron parallel velocity. The magnitude of the electron temperature gradient, on the other hand, can modify significantly the value of the GAM frequency as suggested by Eq. \re{Skl1+}, for example, in low beta plasmas or for electromagnetic perturbations with large radial wave lengths.

The electromagnetic character of GAMs can be measured by the radial mode width parameter\cit{SmolyakovNF2010}, $\K = \kr\lambda_{De} c/q\rz\o\sim \zc{}/q\sqrt{\beta}$, which is dependent on several equilibrium quantities ($\Ti$, $\Te$, $\nz$, $q$) and on only one perturbed parameter: the radial wave length ($\lambda_r$). Since the equilibrium quantities can be determined by plasmas diagnostics in general, measurements of ZFs and GAMs frequencies in tokamaks and stellarators may be useful to reveal informations concerning the perturbed radial electric field and, consequently, the mechanism of interaction between drift wave turbulence and zonal flows can better understood. When $\K\rar \infty$ the GAMs may be said to be purely electrostatic and its frequency\cit{SgallaPLA2013} ($\o_{GAM}^{E}$) is independent of $\etae$ but dependents on $\etai$ and on the electron drift frequency, $\ose = -\Te/eBr\Ln$. In the opposite limit, $\K\rar0$, or purely electromagnetic GAMs, the GAM frequency, $\o_{GAM}^{EM} = (3/4 - \etai)^{1/2}\vthi/\rz$, can be considerably smaller and, apparently, unstable if $\etai > 3/4$. However, for the analysis of stability of the low frequency GAMs, it is necessary to take into account the ion parallel motion dynamics on the investigation of this modes. This issue is left for a future paper. In both limits described above the frequency of the GAMs is provided by Eq. \re{Pslim}. For $\K\sim 0.3$, $\K\sim 1$ and $\K\sim 3$, Eqs. \re{Sk1+}--\re{Skl1-} shows approximate expressions for the GAMs frequencies. Accordingly to Eq. \re{Skl1+}, it can be noted that electromagnetic GAMs are eigenmodes candidates and, hence, the possibility to use them in diagnostic applications\cit{ItohPPCF2007} (GAM spectroscopy) to determine the ion temperature radial profile may be explored. 

The highest frequency oscillation with GAMs features observed in TCABR by long distance correlation analysis in biasing experiments\cit{KuznetsovNF2012} was found to be about 20\% higher than the previous theoretical kinetic prediction for the GAM frequency\cit{LebedevPoP1996}. This paper proposes a model that predicts  GAM expressions that are in agreement with the TCABR result for certain values of $\K$, $\K\sim 0.3$ and $\K\sim 3$. In a scenario of small density and temperature gradients ($\Ose/\ogz\sim 0.03$, $\etai\sim \etae\sim 1$) it is expected that the $f\sim 40$ kHz oscillation observed in TCABR is a high frequency electromagnetic GAM with large radial wave length, $\lambda_r\sim 25cm$. The physics of the H mode and the experimental values of the equilibrium quantities at the transport barrier location may add new information to the present model, which needs to be improved in future works to account more accurately for the experimental reality of tokamaks. 

\begin{acknowledgments}
The author acknowledges with gratitude helpful discussions with A. I. Smolyakov and A. G. Elfimov. This work was supported by CNPq (National Council of Brazil for Science and Technology Development) and by the NSERC Canada (Natural Sciences and Engineering Research Council of Canada).
\end{acknowledgments}
\appendix
\section{Useful relations}
\l{A}

The expressions for the $\eb\t\bb$ flow in a plasma and its divergence (derived from basic vector relations) are generically given by
\bel{A1}
\vb_E = \f{\bh\t\nb\Phi}{B}\co \bh = \f{\bb}{B}\co
\ee
\bel{A2}
\di{\vb_E} = - 2\vb_E\cd\nb\lnb.
\ee

For a high aspect ratio tokamak ($\ep = r/\rz\ll1$) with circular magnetic surfaces, the magnetic field can be approximated by $B\approx\bz(1 - \ep\ct)$ and, consequently, it follows that
\bel{B1}
\nb\lnb = \f{-\ct\er + \st\eth}{\rz}\co \bh\cd\nb\lnb = \np\lnb =  \f{\ep}{\rz}\st.
\ee

Therefore, considering \re{B1}, the perturbed contribution of the $\eb\t\bb$ flow is
\bel{B2}
\vbE = \f{\vthi}{2}\bp{-\f{\rv{i}}{r}\dptheta{\Pbc{}}\er + i\kr\rv{i}\Pbc{}\eth}
\ee
and, if $\kr\gg r^{-1}$ and $\zc{}\ll1$, its divergency can be approximated to
\bel{B2a}
\di\vbE\approx -i\f{\vthi}{\rz}\Pbc{0}\zc{}\st
\ee

With respect to the parallel velocity, $\vp = \vb\cd\bb/B$, it follows that 
\bel{A3}
\di{(\vp\bh)} = \np\vp -  \vp\np\lnb\approx \inv{q\rz}\dptheta{\vp}
\ee
and, since $\vb\approx \vb_E + \vp\bh$ in the MHD order ($|\vbE|\sim \vthi$), 
\bel{A3a}
\bh\bh:\nb\vb = \bh\bh:(\nb\vb)^T = -\vb_E\cd\nb\lnb + \np\vp.
\ee

The parallel viscosity tensor can be expressed as\cit{SmolyakovNF2009}
\bel{A4}
\boldsymbol{\pi}_\parallel = \f32\pi_\parallel\bp{\bh\bh - \f{\it}{3}}\co \pi_\parallel = \f{2}{3}(p_\parallel - p_\perp)\co 
\ee
and, consequently, by considering the approximations mentioned above, it follows that
\bel{A5}
\bh\t\di{\boldsymbol{\pi}_\parallel} =  \f 32\pi_\parallel\bh\t\bh\cd\nb\bh - \inv2\bh\t\nb\pi_\parallel\co
\ee 
\bel{A5a}
\bh\cd\di{\boldsymbol{\pi}_\parallel} = \np\pi_\parallel + \f32\pi_\parallel\np\lnb.
\ee

From the components of the velocities obtained from the momentum equation, see eqs. \re{f2} and \re{fvel}, the respective contributions of the current density ($\jb_\alpha = \es \ns\vb_\alpha$) can be computed from the following expressions:
\bel{A6}
\jbc p = \f{\bh\t\nb p}{B}\co \jbc\pi = \f{\bh\t\di{\boldsymbol{\pi}_\parallel}}{B}\co \jbc I = \f{mn}{B}\bh\t\dt{\vb_E}
\ee
where the perturbed contributions can be approximated to
\bel{B3}
\jbtc{p}^{(\alpha)} = \f{e\nz\vthi}{2}\bp{- \f{\rv{i}}{r}\dptheta{\pcc{\alpha}{}}\er + i\kr\rv{i}\pcc{\alpha}{}\eth }\co \pcc{\alpha}{} = \f{\pth{\alpha}}{\nz\Ti}
\ee
\bel{B4}
\jbtc{\pi}^{(i)} = \f{e\nz\vthi}{4}\bc{ \rv{i}\bp{-\f{3}{\rz}\picc{i}{}\st  + \inv{r}\dptheta{\picc{i}{}}}\er - i\kr\rv{i}\picc{i}{}\eth }\co \picc{i}{} = \f{\tilde{\pi}^{(i)}}{\nz\Ti}
\ee
\bel{B5}
\jbtc{I}^{(i)} = \f{e\nz\vthi}{2}\f{\rv{i}}{\rz}\f{\O}{q}\kr\rv{i}\Pbc{0}\er\co \Pbc{0} = \f{e\Phitc{0}}{\Ti}.
\ee
The divergence of these quantities can be computed from vector relations and similarly to \re{A2} and \re{B2a} yields the following results:
\bel{A7}
\di{\jbtc{p}^{(\alpha)}} = -2\jbtc{p}^{(\alpha)}\cd\nb\lnb\approx -\f{e\nz\vthi}{\rz}\bp{\f{\rv{i}}{r}\dptheta{\pcc{\alpha}{}}\ct + i\zc{}\dptheta{\pcc{\alpha}{}}\st}
\ee
\bel{A8}
\di{\jbtc{\pi}^{(i)}} = \jbtc{\pi}^{(i)}\cd\nb\lnb\approx - \f{e\nz\vthi}{4\rz}\bc{ \f{\rv{i}}{r}\dptheta{\picc{i}{}}\ct +  i\kr\rv{i}\picc{i}{}\st}\co 
\ee
\bel{A9}
\di{\jbtc{I}^{(i)}} = \jbtc{I}^{(i)}\cd\nb\ln(n/B^2) - \f{mn}{B^2}\dpt{}\di{\nb_\perp\Phitc{}}\approx - \f{e\nz\vthi}{2\rz}\bp{i\zc{} - \f{\rv{i}}{\Ln}}\f{\O}{q}\zc{}\Pbc{0}.
\ee

When the perturbed magnetic field (mainly in the perpendicular direction) is considered, in terms of the parallel vector potential $\Apt{}$, the following approximation can be used:
\bel{A10}
\bbt = \curl{(\Apt\bh)} = -\bh\t\nb\Apt + \Apt(\curl{\bh})\approx \inv{r}\dptheta{\Apt}\er - i\kr\Apt\eth
\ee
where 
\bel{A10a}
\curl\bh = \underbrace{\muz\jb/B}_{\Or{\beta/\Ln}} + \underbrace{\bh\t\nb\lnb}_{\Or{1/\rz}}\co \dpr{\Apt}\rar i\kr\Apt.
\ee
Note that, although  $\kr\gg r^{-1}$ is considered, we need to keep the radial component of $\bbt$ in \re{A10} to account for the coupling with equilibrium quantities (density and pressure).

In view of eq. \re{A10}, the parallel current density can be computed as follows:
\bel{A12}
\tilde{j}_\parallel = \f{\bh\cd(\curl\bbt)}{\muz} = \inv{\muz r}\bc{\dpr{(r\tilde{B}_\theta)} - \dptheta{\tilde{B}_r}}\approx \f{\kr^2}{\muz}\Apt
\ee
and, similar to eq. \re{A3}, the correspondent divergence can be approximated to
\bel{A13}
\di{(\tilde{j}_\parallel\bh)} = \f{\kr^2}{\muz q\rz}\dptheta{\Apt} = \f{e\nz\vthi}{\rz}\f{\K^2\O^2}{q\te}\dptheta{\Abc{}}\co \Abc{} = \f{e\vthi\Apt}{\Ti}.
\ee

%

\end{document}